\documentstyle{lamuphys}
\makeatletter
\let\chapter\hid@chapter
\makeatother
\input{psfig.tex}
\begin{document}
\pagenumbering{arabic}
\title{X-ray Variability \index{variability!X-ray} in M87: 1992 - 1998}
\author{D. E. Harris\inst{1}, J. A. Biretta\inst{2}, and
W. Junor\inst{3}}
 
\institute{Smithsonian Astrophysical Observatory, 60 Garden St,
Cambridge MA 02138, USA
\and
Space Telescope Science Institute, 3700 San Martin Dr., Baltimore, MD,
21218 USA
\and
University of New Mexico, 800 Yale Blvd. NE, Albuquerque, NM, 87131
USA}
 
\maketitle

\begin{abstract}
Beginning in 1995 June, we have obtained an observation of M87 with
the ROSAT High Resolution Imager (HRI) \index{ROSAT!HRI} every 6
months.  We present the measurements of X-ray intensity for the core
\index{X-ray!nucleus} and knot A \index{X-ray!knot A} through 1998
January.  We find significant changes in both components.  For the
core, intensities measured in 95 Jun, 96 Dec, and 97 Dec are roughly
30\% higher than values obtained at three intervening times.  For knot
A, a secular decrease of approximately 15\% is interrupted only by an
intensity jump (3~$\sigma$) in 1997 Dec.  Because the background used
for subtraction is probably underestimated, we suspect the actual
variation is somewhat greater than these values indicate.
\end{abstract}
\section{Introduction}

The initial results of our X-ray monitoring \index{X-ray!monitoring}
of M87 (\cite{hbj}, HBJ hereafter) were based on a re-analysis of two
Einstein Observatory \index{Einstein Observatory} HRI observations, a
ROSAT archival HRI pointing in 1992 Jun, and the first two of our
ongoing series of an observation every 6 months (1995 Jun and Dec).
There we demonstrated that the core was variable at the 15\% level and
that knot A showed a gradual decline in intensity.  In this paper we
include data from 4 additional observations, with the most recent
consisting of two segments: 1997 Dec 14 (8 ks) and 1998 Jan 05 (20.6
ks).  We have also re-measured some of the earlier data to ensure
that all data were treated uniformly.

\section{Review of Problems and Measuring Techniques}
 
There are three known problems affecting accurate photometry of ROSAT
HRI data.  The first is that the core and knot A (the two brightest
features which we can measure with a 30 ks observation) are separated
by only 12$^{\prime\prime}$, thus precluding the use of the standard
HRI aperture with r=10$^{\prime\prime}$.  The second problem is that
the central region of M87 is X-ray bright with a complex emission
distribution (\cite{hbjring}), thereby making it difficult to estimate
the surface brightness that would have been observed at the locations
of the core and knot A if they were absent.
 
The most serious problem however, is the image degradation which means
that the effective point response function (PRF) is widened and
distorted from its quasi-Gaussian shape (FWHM $\approx$
5.5$^{\prime\prime}$) to an irregular distribution with FWHM of
7$^{\prime\prime}$ to 10$^{\prime\prime}$.  The degradation is caused
by bad aspect solutions; sometimes associated with the spacecraft
wobble and occasionally to time segments for which the aspect solution
can be up to 10$^{\prime\prime}$ (or more) away from the primary
solution.  This situation means that we have no assurance that our
standardized measuring areas contain the same fraction of total source
counts for each observation.
 
To mitigate the severity of these problems, we made two separate
measurements.  For the first, we take the ratio of net counts in
circular apertures (r=6$^{\prime\prime}$) centered on the core and
knot A.  To first order, we expect that the effective PRF will be the
same for these close sources.  Consequently the major error of this
approach is that we will measure different fractions of the total
source counts in different observations.  Since the background
correction uses a region with a slightly lower surface brightness than
that found adjacent to the core and knot A, the non-variable
background component will increasingly dilute any variable feature as
the effective PRF gets larger; i.e. fewer photons from the unresolved
component will remain within the small measuring aperture.
 
The second measurement utilizes two adjacent rectangles rotated by
20$^{\circ}$.  Each rectangle is 16$^{\prime\prime}$ by
26$^{\prime\prime}$ with their common border (26$^{\prime\prime}$)
perpendicular to the line joining the core and knot A, and centered on
the `reference point'.  The reference point is halfway between the
core and knot A.  Surrounding these two rectangles is the background
`frame' which is 10$^{\prime\prime}$ wide.  This background frame is
used both for the circular and rectangular measurements.  The geometry
is depicted in figure 2 of HBJ.  The larger area of the rectangle
compared to the small circle is designed to include most of the source
counts even when there is substantial degradation of the PRF.
However, since it will include more of the non-variable background,
the actual variability should be somewhat greater than that found by
this technique.
 
As a control, we also measure the net counts in a circular aperture
(r=12$^{\prime\prime}$) centered 45$^{\prime\prime}$ to the SE of the
reference point (PA=110$^{\circ}$).  This location is on a plateau in
the X-ray brightness distribution.  Since we use the same background
frame as for the other measurements, the resulting value is always
negative.
 
Finally, to accommodate any changes in quantum efficiency we employed a
`self-calibration' by measuring the count rate of the bright part of
the cluster gas within a large circle with radius
276$^{\prime\prime}$, but excluding the adjacent rectangles containing
the core and knot A.  For this measurement, we used a background
annulus with radii of 280$^{\prime\prime}$ and 300$^{\prime\prime}$.
All measured net count rates were then multiplied by a correction
factor (always 5\% or less) so that the cluster gas net count rate
would be the same as that for our 'fiducial' observation of 1995 June.

\begin{figure}
\centerline{
\psfig{figure=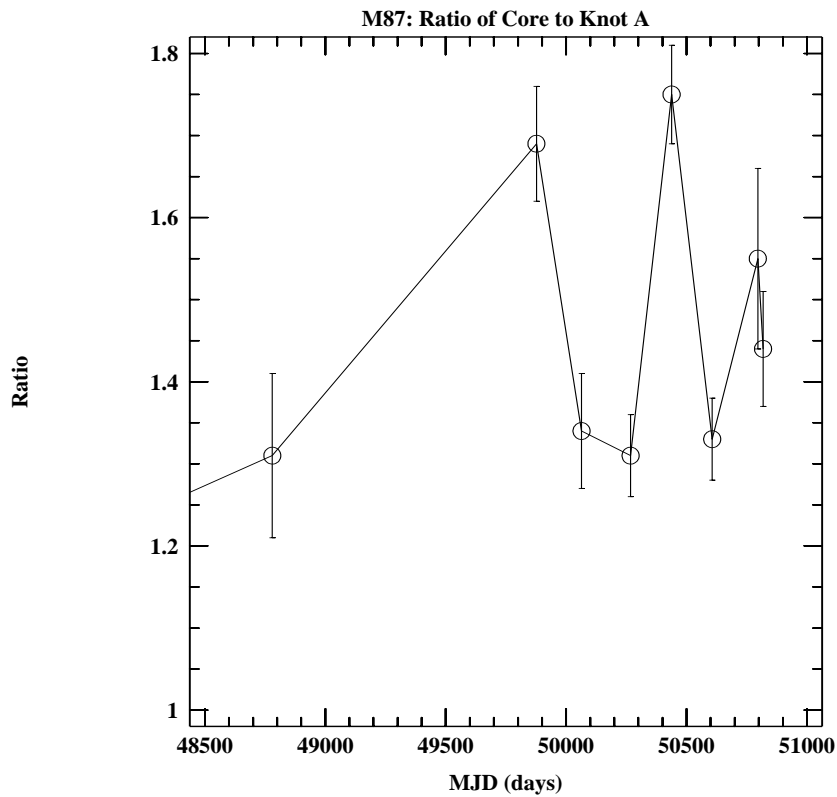,width=15cm,angle=-90}
}
\caption{\label{fig1}The ratio of the core to knot A.  Values are the
ratio of net counts as measured in circular apertures with radius of
6$^{\prime\prime}$.}
\end{figure}

\section{Results}
 
The measured values are shown in figures 1 (the ratios) and 2 (the
count rates).  It can be seen that the major features of the ratio
plot mimic the count rate variability for the core.  This is
consistent with expectations that the core will be more variable than
knot A.  The decrease in the intensity of the core following the
observation of 95 Jun (MJD = 49876) coincides with the decrease
observed with the HST between MJD 49840 and MJD 49921 (see figure 3 of
\cite{zlatko}).

\begin{figure}
\centerline{
\psfig{figure=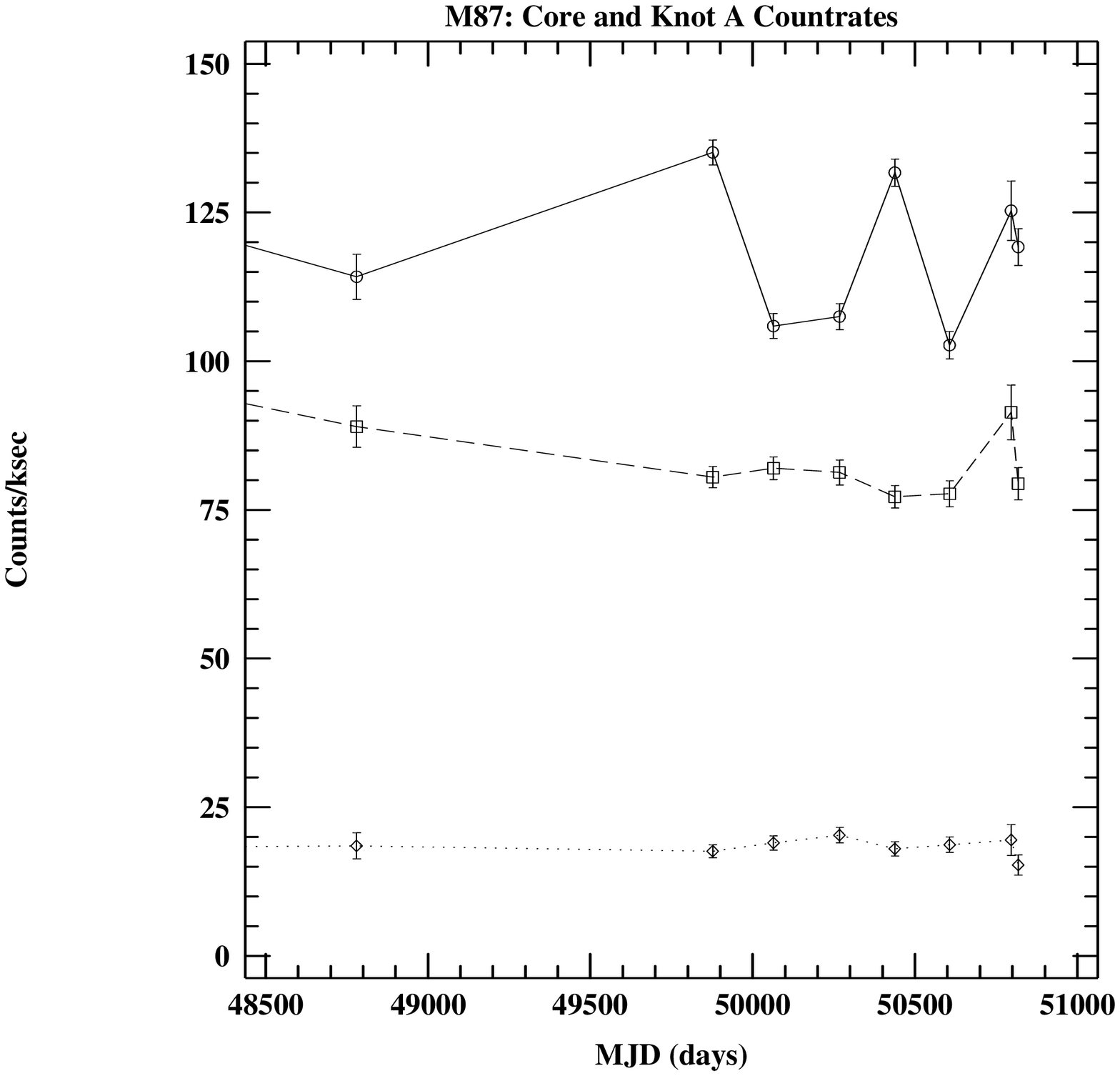,width=15cm,angle=-90}
}
\caption{\label{fig2}The net count rates for the core, knot A, and a
control region.  The solid line is for the core, the dashed line is
for knot A, and the control circle values are connected with a dotted
line.  Note that the control region is the net counts in a circle
45$^{\prime\prime}$ SE of the reference point, with the background
from the frame.  Therefore all control values are negative, but are
plotted here as positive numbers.  Thus the small apparent drop for
the last measurement is actually in the opposite sense from those of
the core and knot A.  The lines running off the left side of the
figure connect to the Einstein values (HBJ).}
\end{figure}

In \cite{bsh}, we analyzed the Einstein \index{Einstein Observatory}
HRI data and argued that the X-ray emission from knot A was most
likely synchrotron emission \index{emission!synchrotron} rather than
thermal bremsstrahlung \index{emission!thermal bremsstrahlung} or
inverse Compton \index{emission!inverse Compton} emission.  For a
magnetic field \index{magnetic field strength} strength of $\approx$
200 $\mu$G (the minimum pressure field), we found that electrons with
Lorentz energy factors, $\gamma$, of 2$\times$10$^{7}$ were required
and typical halflives were of order ten years.  The observed decrease
of knot A (Fig.~\ref{fig2}) is consistent with these estimates, but we
do not yet understand why only a small fraction of shocks
\index{acceleration!shocks} produce enough $\gamma~=~10^{7}$ electrons
to generate an X-ray intensity detectable with current technology.  We
suspect that even the knot A shock ($\approx$ 70 pc across as measured
at radio and optical wavelengths) is not a uniform single entity, but
may well display a complex brightness distribution at the highest
energies where synchrotron losses are severe.  Thus we expect that
higher resolution X-ray observations will show occasional bright,
compact components which will not persist many years
(e.g. \cite{biretta} find optical features with these
characteristics).  If the 1997 Dec increase in knot A is real, it
could represent emission from such a feature.  AXAF \index{AXAF}
observations have been proposed to obtain 8 monthly exposures with the
High Resolution Camera.

%
%

\end{document}